\documentclass[conference,10pt]{IEEEtran}

\usepackage{epsfig,rotating,setspace,latexsym,amsmath,epsf,amssymb,amsfonts,bm,theorem,subfigure,epstopdf}

\DeclareMathOperator*{\argmax}{arg\,max}

\usepackage{cite,authblk}

\usepackage{bbm}

\usepackage[ruled,vlined]{algorithm2e}

\usepackage{color}

\IEEEoverridecommandlockouts
\allowdisplaybreaks

\begin{document}
\bstctlcite{IEEEexample:BSTcontrol} 

\title{Autoencoder-based Communications with Reconfigurable Intelligent Surfaces\thanks{This effort is supported by the U.S. Army Research Office under contract W911NF-21-C-0015. The content of the information does not necessarily reflect the position or the policy of the U.S. Government, and no official endorsement should be inferred.}}

\author[1]{Tugba Erpek}
\author[1]{Yalin E. Sagduyu}
\author[2]{Ahmed Alkhateeb}
\author[3]{Aylin Yener}

\affil[1]{\normalsize Intelligent Automation, A BlueHalo Company, Rockville, MD 20855, USA
}
\affil[2]{\normalsize Arizona State University, Tempe, AZ 85281, USA}
\affil[3]{\normalsize The Ohio State University, Columbus, OH 43210, USA
}

\maketitle
\begin{abstract}
This paper presents a novel approach for the joint design of a reconfigurable intelligent surface (RIS) and a transmitter-receiver pair that are trained together as a set of deep neural networks (DNNs) to optimize the end-to-end communication performance at the receiver. The RIS is a software-defined array of unit cells that can be controlled in terms of the scattering and reflection profiles to focus the incoming signals from the transmitter to the receiver. The benefit of the RIS is to improve the coverage and spectral efficiency for wireless communications by overcoming physical obstructions of the line-of-sight (LoS) links. The selection process of the RIS beam codeword (out of a pre-defined codebook) is formulated as a DNN, while the operations of the transmitter-receiver pair are modeled as two DNNs, one for the encoder (at the transmitter) and the other one for the decoder (at the receiver) of an autoencoder, by accounting for channel effects including those induced by the RIS in between. The underlying DNNs are jointly trained to minimize the symbol error rate at the receiver. Numerical results show that the proposed design achieves major gains in error performance with respect to various baseline schemes, where no RIS is used or the selection of the RIS beam is separated from the design of the transmitter-receiver pair. 

\end{abstract}

\section{Introduction}\label{sec:Introduction}
Deep learning provides powerful means for wireless systems to learn from rich representations of spectrum data and solve complex tasks that are hard (if not infeasible) to solve by analytical methods or expert knowledge \cite{bcDL}. One particular application of deep learning in the wireless domain is the physical layer design that ranges from the optimization of individual communication blocks (such as beamforming) to the joint design of the entire transmitter-receiver chain. By accounting for channel effects, examples of deep learning for optimization of individual communication blocks include channel decoding \cite{decoding}, channel estimation \cite{estimation}, channel access \cite{deepwifi}, power control \cite{kimpower}, beam selection for multiple-input multiple-output (MIMO) communications \cite{MIMO}, and beam prediction for initial access in directional transmissions \cite{IA}.

In conventional communication systems, the operations at the transmitter and the receiver are divided into blocks following a modular design approach. For instance, the modulation block maps symbols to complex numbers, namely (I/Q) samples, at the transmitter, whereas the demodulation block maps the received signals back to symbols. Similarly, channel coding/decoding corresponds to another set of blocks. While this modular design has been considered the building block of current communication systems, it does not necessarily lead to the optimal design of the entire transmitter-receiver chain. The reason is two-fold. First, the transmitter and the receiver chains (and even individual blocks within each chain) are designed and optimized only separately in conventional communication systems. Second, the blocks in conventional communication systems are typically fixed and their design does not adapt to channel conditions. For example, modulation schemes such as binary phase shift keying (BPSK) or quadrature phase shift keying (QPSK) correspond to fixed constellations. While it is possible to select different constellation sizes over time (through adaptive modulation \cite{AMC}) adapting to channel conditions, the set of the underlying constellations is pre-determined and these constellations are not designed with respect to channel conditions. 

\begin{figure}[h]
	\centerline{\includegraphics[width=\linewidth]{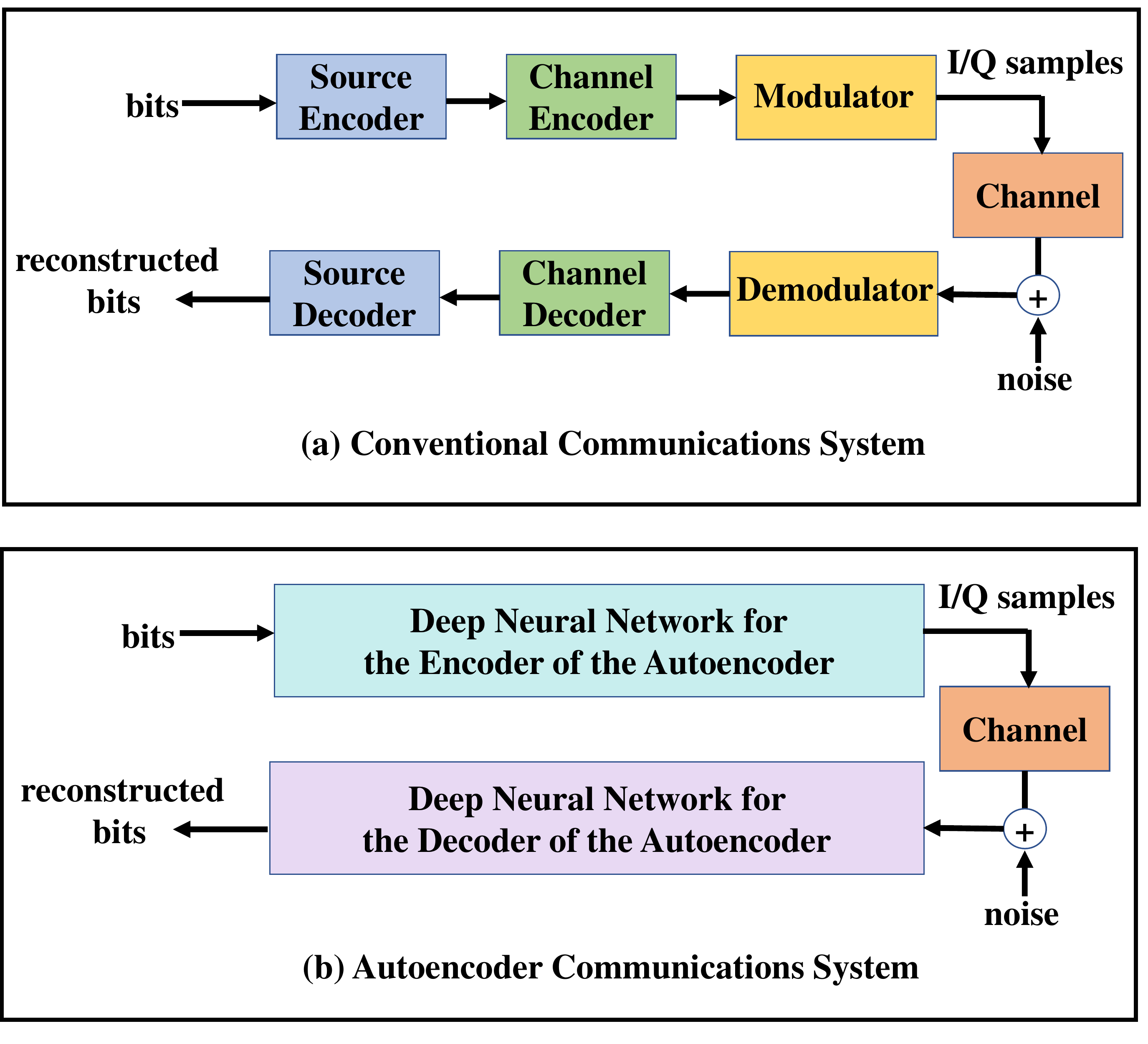}}
	\caption{Conventional vs. AE communication system.}
	\label{fig:AE}
\end{figure}

The high data rate and reliability needs of next generation (nextG) communication systems necessitate a joint design of the transmitter and receiver chains by adapting to the channel data, moving away from the modular design paradigm. For this joint design, end-to-end learning of communication systems using autoencoders (AEs) was introduced in \cite{intromlcomsys}. The transmitter and the receiver are formulated as the encoder and decoder of an AE system  \cite{goodfellow2016deep}, respectively, as shown in Fig.~\ref{fig:AE}. Here, the encoder and the decoder are two (different) deep neural networks that are trained jointly to optimize the end-to-end communication performance such as minimizing the error rate at the receiver. 

This AE system is optimized using a channel model between the encoder and the decoder as opposed to block-wise optimization as in conventional communication systems. It was shown in \cite{intromlcomsys} that AEs for single antenna communications can match and exceed the performance of near-optimal existing baseline modulation and coding schemes. This AE approach was shown in \cite{dlota} to work effectively over the air using unsynchronized off-the-shelf software-defined radios (SDRs), where transfer learning is used to overcome the mismatch between the stochastic channel model used during training and the actual channel observed in test time. AE-based techniques were extended in \cite{FelixOFDM},\cite{KimOFDMPAPRDL} to Orthogonal Frequency Divison Multiplexing (OFDM). In \cite{timMIMO}, the preliminary results of the AEs were provided for $2\times2$ spatial diversity and spatial multiplexing MIMO systems. AEs were also used  for the interference channel setting in \cite{erpekICC}, which showed that the interference can be successfully learned and cancelled at the receivers when two AE systems (transmitter and receiver pairs) are jointly trained. Beyond Additive White Gaussian Noise (AWGN) channels, AE communications was also extended in \cite{fading} to frequency- and time-selective fading channels. 

\begin{figure}
	\centerline{\includegraphics[width=1\linewidth]{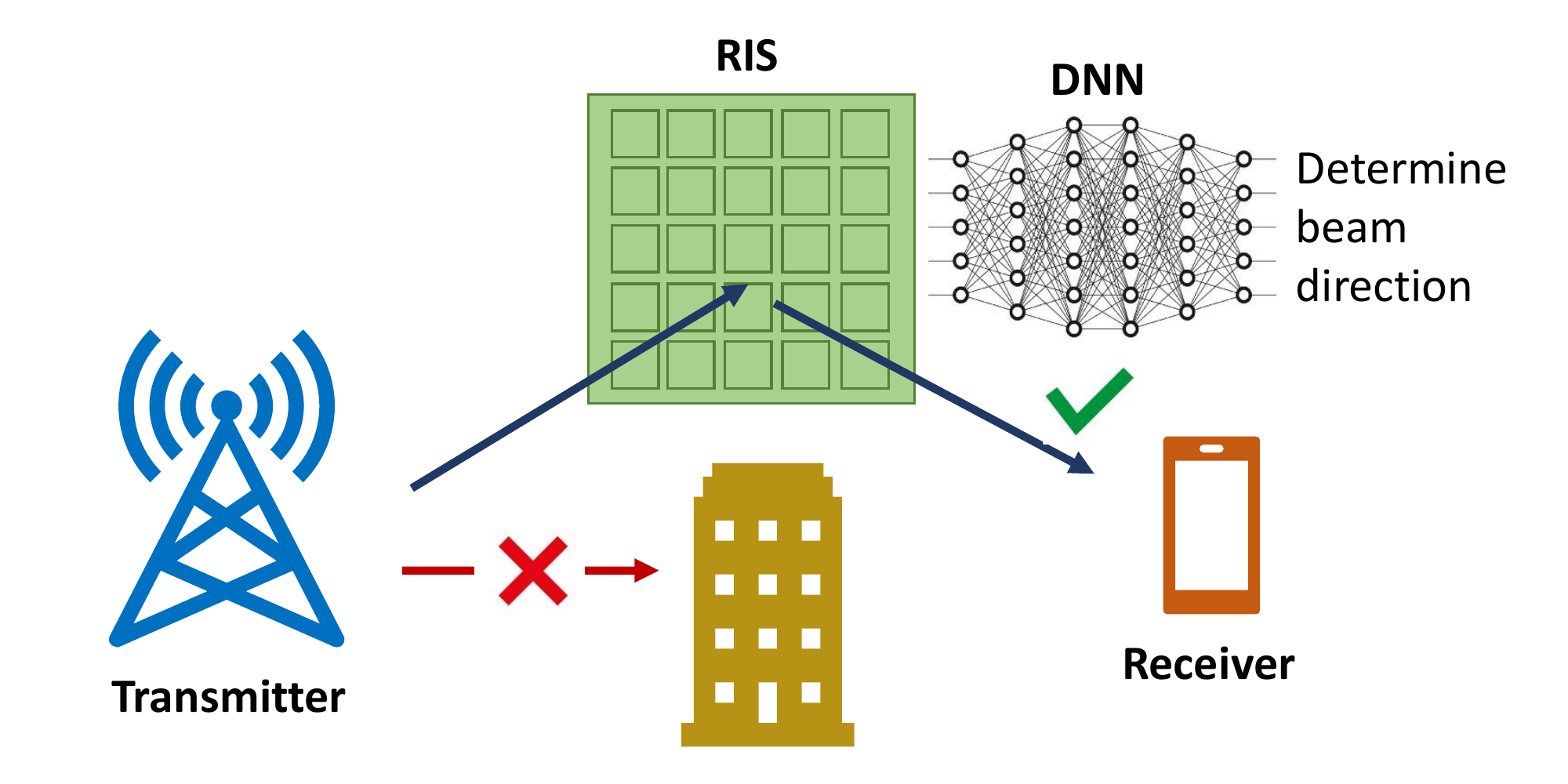}}
	\caption{RIS-aided communications.}
	\label{fig:RIS}
\end{figure}

Another emerging design aspect in wireless communications is the use of reconfigurable intelligent surfaces (RISs) to mitigate the effects of path loss and multipath attenuation. One scenario where RIS can be useful is boosting communications in mmWave bands that are vulnerable to physical obstructions. RIS is a programmable antenna array placed in the communication medium that can be used to beamform the received signal towards the direction of interest \cite{Basar, Huang, Han19, LiRIS, peiRIS, RISBjornson, Nof, Taha}. RIS has benefits particularly when there is no direct link between the transmitter and the receiver. In this scenario, an RIS can be placed between the transmitter and the receiver to focus the transmitted signals toward the receiver, as shown in Fig.~\ref{fig:RIS}. It was shown in \cite{Trichopoulos} that when both the transmitter and receiver employ directional antennas and the distances between the transmitter-RIS and RIS-receiver are $5$m and $10$m, respectively, the RIS achieves $15$-$20$dB gain in the signal-to-noise ratio (SNR) at the receiver.  

In this paper, we focus on a communication system with a transmitter, an RIS, and a receiver. We combine the AE-based communication system design with the RIS beamforming vector (codeword) selection to optimize the transmitter and receiver functionalities by taking the channel effects and the RIS directionality into account. The transmitter, the receiver and the RIS operations are modeled as individual DNNs that are jointly trained to minimize the symbol error rate (SER) at the receiver. This approach accounts for the channel effects including those induced by the RIS and configures the transmitter, receiver and RIS operations with respect to channel conditions. We show that this joint optimization approach reduces the SER significantly relative to the case when no RIS is used, in the presence of physical obstructions. We highlight that the RIS beam index needs to be selected jointly with the optimization of transmitter-receiver functionalities for best performance. 

The remainder of the paper is organized as follows. Section \ref{sec:SystemModel} describes the system model. Section \ref{sec:Design} presents the joint design of the transmitter-receiver pair and the RIS. Section \ref{sec:Performance} reports the performance evaluation results. Section \ref{sec:Conclusion} concludes the paper.  

\section{System Model}\label{sec:SystemModel}
\begin{figure}[h]
	\centerline{\includegraphics[width=1\linewidth]{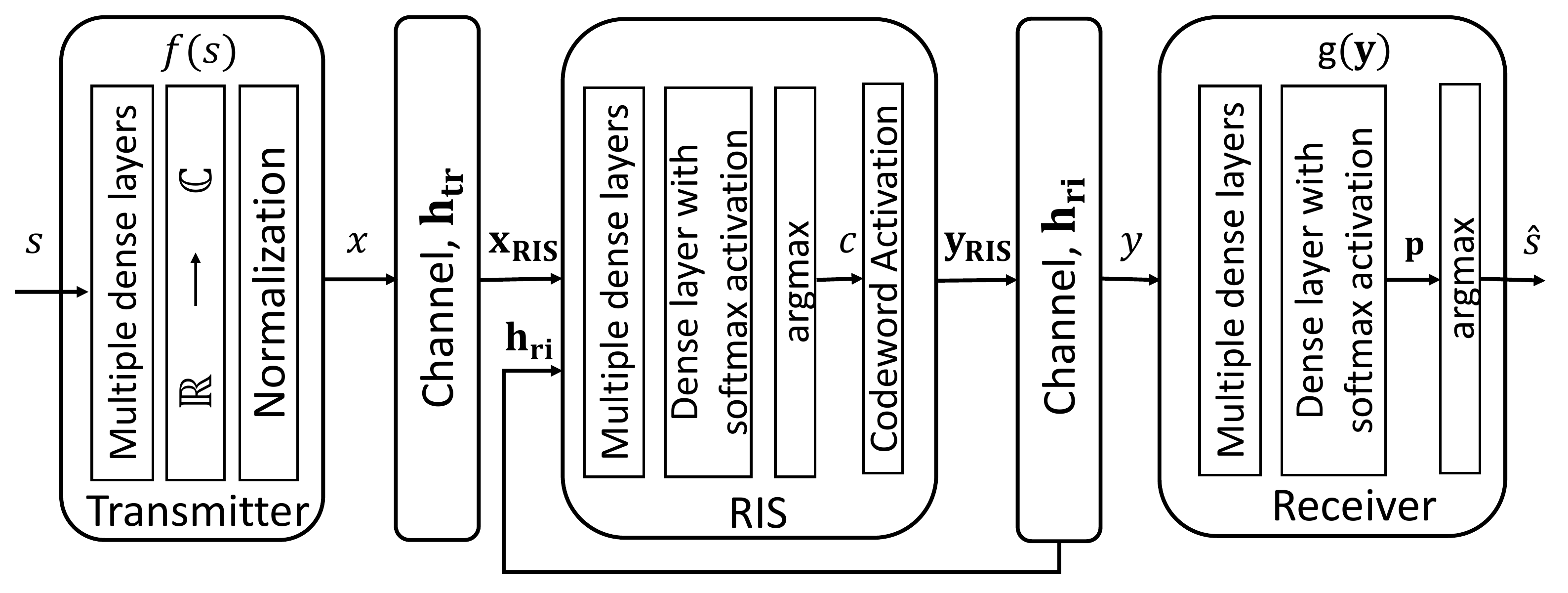}}
	\caption{The system model for RIS-aided AE-based wireless communications. As the encoder, the transmitter takes $s$ as input and generates modulated symbols $x$ where $x=f(s)$. Similarly, as the decoder, the receiver takes $y$ as the input and maps those values to symbol estimates $\hat{s}$ where $\hat{s}=g(y)$.}
	\label{fig:AE_TX_RX}
\end{figure}

We consider an RIS-aided AE-based communication system where a transmitter aims to communicate with a receiver. There is no direct link between them, e.g., due to a physical obstruction. An RIS is placed between the transmitter and the receiver to enable communications from the transmitter to the receiver, as shown in Fig.~\ref{fig:RIS}. We assume that single omni-directional antennas are used at the transmitter and the receiver. The signal transmitted from the transmitter is passively reflected at the RIS towards the direction of the receiver. The system model is shown in Fig.~\ref{fig:AE_TX_RX}.

Two separate deep neural networks (DNN) are used to represent the transmitter and receiver operations. The DNN at the transmitter corresponds to an encoder and the DNN at the receiver corresponds to a decoder. This encoder-decoder pair forms an AE system that is trained to match the input of the encoder with the output of the decoder. The transmitter takes the symbols $s$ as the input and transmits the modulated complex symbol $x$ as the output. The intended receiver uses a pre-trained DNN classifier to detect the ongoing signal that is reflected by an RIS. The RIS is equipped with $N$ reconfigurable antenna elements. When the transmitter sends $x$, the signal at the RIS antennas is given by
\begin{equation}\label{eq:ris}
	\mathbf{x_{ris}}(x) = \mathbf{h_{tr}}x, 
\end{equation}
where $\mathbf{h_{tr}}\in \mathbb{C}^{N \times 1}$ is the channel between the transmitter and the RIS. We assume that the phase shift of the RIS element is quantized and represented with 1 bit where each RIS element introduces either $0^{\circ}$ or $180^{\circ}$ phase shift and $\kappa$ insertion loss to the signal. Thus, the signal at the output of the RIS is given by
\begin{equation}\label{eq:output}
	[\mathbf{y_{ris}}(x)]_{n} = \psi_{n}[\kappa\mathbf{x_{ris}}(x)]_{n},  \;\;\;\; n = 1,\cdots, N, 
\end{equation}
where $\psi_{n}\in \{-1,1\}$ or $\psi_{n}=e^{j\theta_n}$ and $\theta_n$ corresponds to the phase shifts ($\theta_{n} \in \{0,\pi\}$). The RIS is also represented with a DNN. It takes $\mathbf{x_{ris}}(x)$ and the channel between the RIS and the intended receiver, $\mathbf{h_{ri}}\in \mathbb{C}^{N \times 1}$, as an input and returns the index of the codeword that maximizes the SNR at the receiver as the output. There is no added noise at the RIS since it is a passive device. The received signal at the intended receiver is given by
\begin{equation}\label{eq:receive}
	y(x) = \mathbf{h_{ri}}^{T}\mathbf{y_{ris}}(x)+n_r,
\end{equation}
where $n_r$ is the AWGN at the intended receiver. This channel formulation takes the channel gain and the phase shift from each RIS element to the intended receiver into account.


\section{Joint Optimization of the Transmitter-Receiver Pair and the RIS} \label{sec:Design}
\subsection{Transmitter Design} \label{subsec:TX}
The encoder at the transmitter is designed using a feed-forward neural network (FNN) architecture. The input symbols go through dense layers. The input to the transmitter consists of symbols $s$. The transmitter communicates one out of $2^k$ possible messages to the receiver, where $k$ is the number of information bits in each symbol. Note that we consider $k=2$ in performance evaluation. The output of the last dense layer is reshaped to generate complex numbers as the output; i.e., complex numbers with even indices as the real part and odd indices as the imaginary part. 

We impose an average power constraint on the transmitter. For that purpose, the normalization layer is used to normalize the transmitter output so that the average power constraint is satisfied; i.e., $E[\mathbf{x^*}\mathbf{x}]\leq P$. Note that the normalization layer is designed as a custom layer and can be regarded as a neural network layer without any trainable parameters. The averaging is applied over mini-batch. The normalized output can be written as 
\begin{equation}
x = \frac{x'}{\sqrt{\frac{1}{M}\sum_{i=1}^M \lVert \mathbf{x'}_i \rVert_2^2}},
\end{equation}
where $x'$ is the output of the real-to-complex layer and $M$ is the mini-batch size.
 
The transmitter output, $x$ can be thought as modulated symbols as in the conventional communication systems. Instead of using a known constellation scheme with linear decision regions such as BPSK for $k=1$ or QPSK for $k=2$, the optimal constellation points are learned by the AE system based on the channel characteristics. The output constellation diagram may include nonlinear decision regions as the bit rate increases. 

Table \ref{table:paraTX} lists the number of layers and neurons along with the activation functions used in each dense layer at the transmitter. The transmitter output size is twice the number of inputs with the complex representation of the modulated symbols since I and Q components are used separately.  

\begin{table}
	\caption{The FNN structure at the transmitter.}
	\centering
	{\small
		\begin{tabular}{c|c|c}
			Layers & \# neurons & Activation function\\ \hline \hline
			Input & 1 &    \\ \hline 
			Dense 1 & 256 & ReLU \\ \hline
			Dense 2 & 256 & ReLU \\ \hline
			Output & 2 & Linear \\ \hline
		\end{tabular}
	}
	\label{table:paraTX}
\end{table}

\subsection{Channel Model} \label{subsec:Ch}
We consider a single-path geometric channel model \begin{equation}
\bold{h}=\alpha \: \bold{a}(\theta_l, \phi_l),
\end{equation}
where $\alpha$ is the path loss and $\bold{a}(\theta_l,\phi_l) \in \mathbb{C}^{N \times 1}$ represent the phase shifts at the azimuth and elevation angles of arrival, $\theta_l$ and $\phi_l$, respectively \cite{alkhateeb1}. A phase shift is introduced due to different distances from the transmitter to each antenna element accordingly. The channel remains constant during training. A multiplication layer is built to perform complex multiplications, $\mathbf{h_{tr}}x$ and $\mathbf{h_{ri}}^T\mathbf{y_{ris}}$. The noise layer introduces noise, $n_r$, to the AE system at the intended receiver and it is updated in every training instance. A different noise value is selected with each training example and the noise variance $\sigma$ is adjusted during both training and test times to simulate varying levels of SNR at the receiver. 

\subsection{RIS Codeword Vector Selection}
The RIS codeword vector selection is formulated as an FNN. The signal received at the RIS and also $\mathbf{h_{ri}}$ are used as input to the FNN. The index of the codeword from a pre-defined codebook, $\mathcal{P}$, that maximizes the SNR at the intended receiver is provided as the FNN output. A custom layer is built to activate the RIS antenna elements that correspond to the selected codeword. Table \ref{table:paraRIS} shows the DNN parameters for the RIS codeword selection functionality.

\begin{table}
	\caption{The FNN structure at the RIS.}
	\centering
	{\small
		\begin{tabular}{c|c|c}
			Layers & \# neurons & Activation function\\ \hline \hline
			Input & 66 &    \\ \hline 
			Dense 1 & 400 & ReLU \\ \hline
			Dense 2 & 400 & ReLU \\ \hline
			Dense 3 & 400 & ReLU \\ \hline
			Dense 4 & 400 & ReLU \\ \hline
			Output & 32 & Softmax \\ \hline
		\end{tabular}
	}
	\label{table:paraRIS}
\end{table}

\subsection{Receiver Design} \label{subsec:RX}
The decoder at the receiver is also designed using an FNN architecture. The symbol received at the receiver, $y$, goes through multiple dense layers with the last layer equipped with softmax activation. The receiver input size is $2$ since I and Q components are used separately. Note that the channel estimation and equalization steps are implicitly performed at the receiver. Softmax activation outputs a probability for each symbol. There are a total of $2^{k}$ output classes. The symbol with the maximum probability is chosen as the output. Table \ref{table:paraRX} lists the number of layers and neurons along with the activation functions used in each dense layer at the decoder.

\begin{table}
	\caption{The FNN structure at the receiver.}
	\centering
	{\small
		\begin{tabular}{c|c|c}
			Layers & \# neurons & Activation function\\ \hline \hline
			Input & 2 &    \\ \hline 
			Dense 1 & 1024 & ReLU \\ \hline
			Dense 2 & 1024 & ReLU \\ \hline
			Dense 3 & 1024 & ReLU \\ \hline
			Output & 4 & Softmax \\ \hline
		\end{tabular}
	}
	\label{table:paraRX}
\end{table}

\subsection{The RIS Codebook Vector Design} \label{subsec:Codebook}
The RIS codebook vector is designed to maximize the SNR at the intended receiver. Denoting the RIS reflection vector as $\Psi$, we can write the achievable rate as 
\begin{equation}
	R = \log_2 \left( 1+ \frac{P_t}{\sigma^2} \left| \left( \mathbf{h_{tr}} \odot \mathbf{h_{ri}} \right) ^T \Psi(\phi_n) \right| ^2 \right), \label{eq:R}
\end{equation}
where $P_t$ is the transmit power and $\sigma^2$ is the noise variance. Every element in the $N \times 1$ RIS reflection vector $\Psi$ is implemented using an RF phase shifter. The codebook generation process is described in \cite{Trichopoulos}. 
Assuming phase shifts of the RIS elements are continuous variables, the RIS reflection vector $\Psi^*$ that maximizes the achievable rate $R$ is given by
\begin{align} 
	\Psi^* = &\argmax_{\phi_n \in [0^{\circ},360^{\circ}], \forall n}\log_2 \left( 1+ \frac{P_t}{\sigma^2} \left| \left( \mathbf{h_{tr}} \odot \mathbf{h_{ri}} \right) ^T \Psi(\phi_n) \right| ^2 \right).
\end{align} 

Since we assume that the transmitter-RIS and RIS-receiver communication links are dominated by LoS paths \cite{Trichopoulos}, the RIS reflection vector can be designed as 
\begin{align} 
	\Psi^* = &\argmax_{\phi_n \in [0^{\circ},360^{\circ}], \forall n} \left| \left( \mathbf{h_{tr}} \odot \mathbf{h_{ri}} \right) ^T \Psi(\phi_n) \right| ^2.
\end{align} 

Since we assume LoS channels from the transmitter to the RIS and from RIS to the receiver, the effective channel mainly corresponds to the RIS array response vector for the incident signal direction and the desired reflection direction, as follows.
\begin{align} 
	\Psi^* \hspace{-0.05cm}=\hspace{-0.1cm} &\argmax_{\phi_n \in [0^{\circ},360^{\circ}], \forall n} \hspace{-0.05cm}\left| \left( \mathbf{a_{RIS}^*}(\theta_d, \phi_d) \odot \mathbf{a_{RIS}^*}(\theta_i, \phi_i) \right) ^T \hspace{-0.1cm}\Psi(\phi_n) \right| ^2,
\end{align} 
where $\mathbf{a_{RIS}^*}(\theta, \phi)$ is the RIS array response vector for the angles $\theta$ and $\phi$. Incident and desired reflection directions are $(\theta_i, \phi_i)$ and $(\theta_d, \phi_d)$, respectively. The optimal RIS phase shifting configuration for each element $\psi_n$ in $\mathbf{\Psi}$ is given by
\begin{equation}
    \psi_n = \psi_{i,n}-\psi_{d,n},
\end{equation}
where $\psi_{i,n}$ and $\psi_{d,n}$ are the phase values for the incident and desired reflection direction waves, respectively, on the $n$th RIS unit cell. 

The phase shifts of the array elements of RISs need to be quantized into discrete values due to hardware limitations. We use a 1-bit quantizer and represent the quantized phase shift at the $n$th RIS unit cell as

\begin{equation}
\phi_n^{quant} = \left| 180^\circ \times \text{round} \left(\frac{\phi_n}{180^\circ} \right) \right|.
\end{equation}
As a result, all the phase values in the range $\left[ -90^\circ,+90^\circ \right]$ are rounded to $0^\circ$ and the rest of the values are rounded to $180^\circ$. 

If $\mathcal{P}$ represents a predefined codebook, then our goal is to find $\Psi^*$ in $\mathcal{P}$ that maximizes $R$ at the receiver (given by (\ref{eq:R})) along with optimized modulation and demodulation operations with the AE-based RIS-aided communication system. We limit the choice of the reflection beamforming vector to a predefined codebook $\mathbf{\mathcal{P}}$ with $N$ elements. Each element in the codebook is represented with 1 bit \cite{Trichopoulos}.     

\subsection{Optimization Process} \label{subsec:Opt}

A categorical cross-entropy loss function ($\ell_{CE}$) is used for optimization using gradient descent to select the DNN parameters.  
In this case, $\ell_{CE}$ is given by 

	\begin{equation}
		\ell_{CE}(\boldsymbol{\theta}) = -\frac{1}{M}\sum_{i=1}^{M} \sum_{j=0}^{2^{k}-1} p_{o,j}' \log(p_{o,j}),
	\end{equation}
where $M$ is the mini-batch size, $\boldsymbol{\theta}$ is the set of DNN parameters, $p_{o,i}$ is the softmax layer's  output probability for output class $i$ for observation $o$ and $p_{o,i}'$ is the binary indicator ($0$ or $1$) if class label $i$ is the correct classification for observation $o$. Using a form of stochastic gradient descent (Adam \cite{adam}), weight updates are computed based on the loss gradient using back-propagation.  In this case, we iteratively compute a forward pass, $f(s,\boldsymbol{\theta})$, and a backward pass, 
$\frac{ \partial \ell_{CE} (\boldsymbol{\theta}) }{ \partial \boldsymbol{\theta}}$ and a weight update takes the form of $\Delta w = -\eta \frac{\partial \ell_{CE}(\boldsymbol{\theta})}{\partial \boldsymbol{\theta}}$, where $\eta$ represents a time-varying learning rate. In the backwards pass, the noise layer becomes the identity function (it is used only for forward passes). 

\section{Performance Evaluation}\label{sec:Performance}
\begin{figure} [h]
	\centerline{\includegraphics[width=0.5\linewidth]{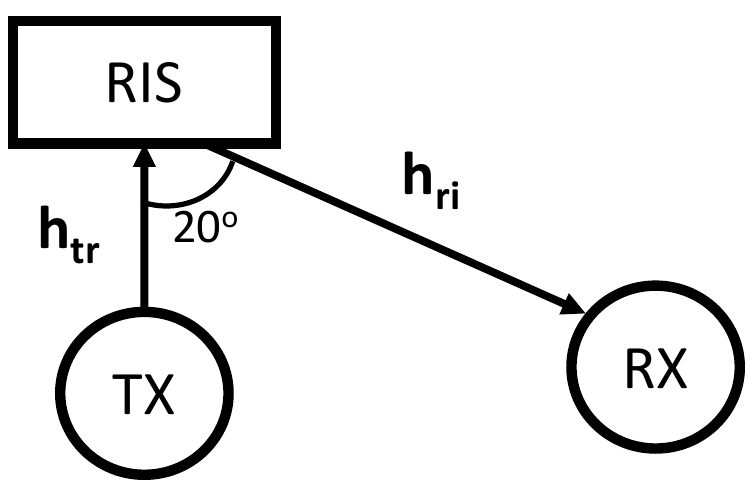}}
	\caption{The topology of the transmitter, the RIS and the receiver used for performance evaluation.}
	\label{fig:scenario}
\end{figure}

The simulation results for both the conventional and AE-based RIS-aided communication systems are provided in this section. The transmitter, the RIS, and the receiver are located as shown in Fig.~\ref{fig:scenario}.

We vary the SNR for the RIS-aided communications (namely the effective SNR from the transmitter to the receiver over the RIS) and measure the SER. Then, we compute the corresponding SNR for the direct link from the transmitter to the receiver by accounting for a loss ($l_o$) due to a physical obstruction between the transmitter and the receiver. 

\subsection{Baseline Communication Scheme}
The SER performance of a QPSK-modulated conventional communication system is simulated  with the AWGN channel as the baseline assuming a direct channel from the transmitter to the receiver.  

\subsection{AE-based RIS-aided communication system}
A set of $\mathcal{P}$ codewords is generated with $\left| \mathcal{P} \right| = 32$. The incident angle is set as $90^{\circ}$ and the designed codebooks direct the signal in reflection angles set between $100^{\circ}$ and $160^{\circ}$. The receiver is placed with a reflection angle of $110^{\circ}$ as shown in Fig.~\ref{fig:scenario}. The RIS antenna spacing is set as $\lambda/2$, where $\lambda$ is the wavelength. The RIS introduces $\kappa_{dB} = 3$ dB insertion loss to the received signal.

In addition to the optimal selection of the RIS beam code (namely, the one that leads to the smallest SER), we also consider benchmark schemes, where we select the RIS beam index randomly or from top-$K$ RIS beam indices (i.e., from the $K$ beam indices that lead to the smallest SERs).  
  
\subsection{Symbol Error Rate Performance}
\begin{figure}
	\centerline{\includegraphics[width=1\linewidth]{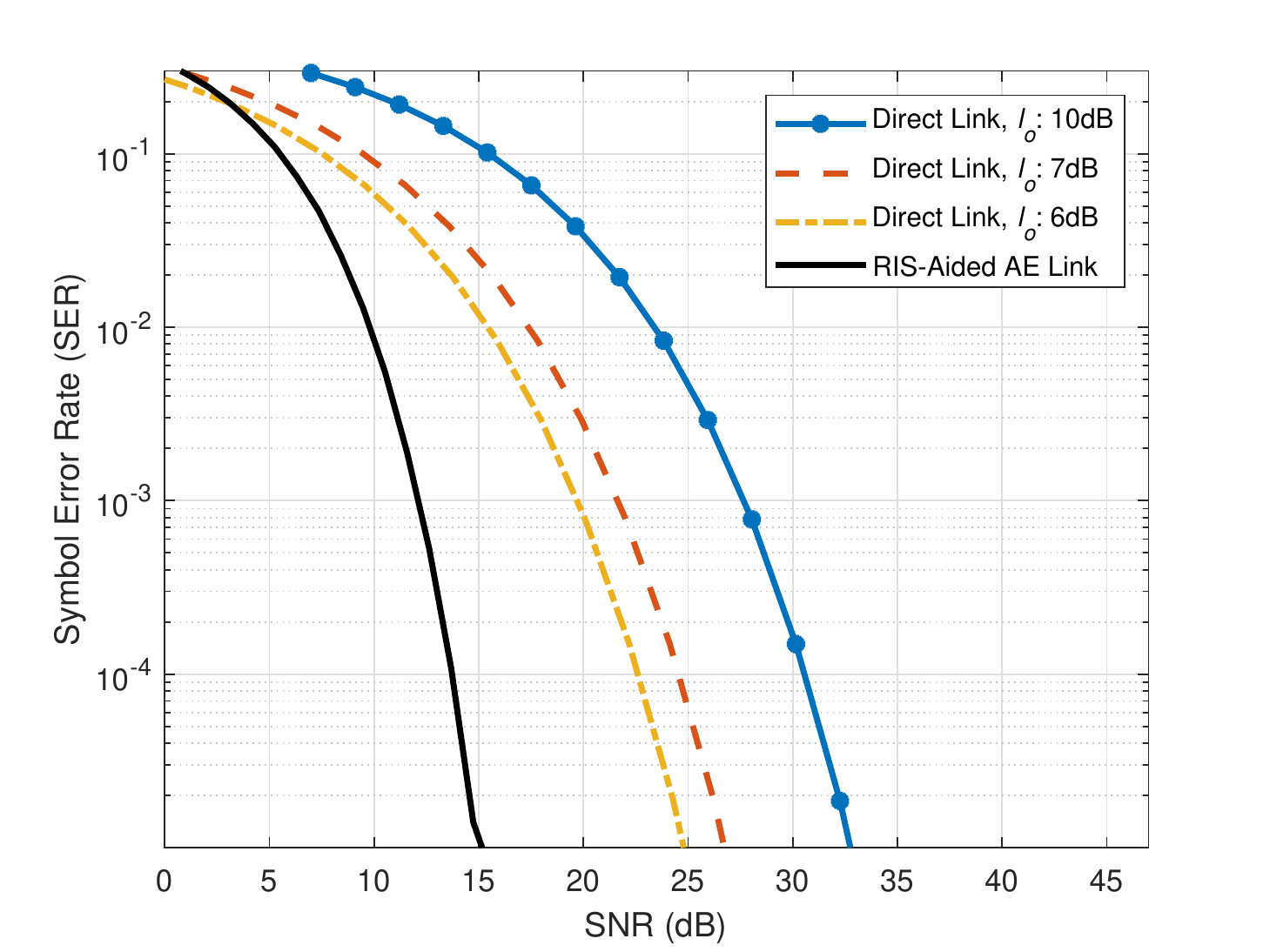}}
	\caption{SER vs. SNR performance for RIS-Aided AE communications and direct communications between the transmitter and the receiver.}
	\label{fig:RISPerf}
\end{figure}

In this section, we compare the performance of the AE-based RIS-aided communications with benchmark schemes. The first scheme is based on using the direct link from the transmitter to the receiver without utilizing the RIS. This scheme is hindered by the physical obstruction between the transmitter and receiver that we model as an obstruction loss, $l_o$. The SER is shown in Fig.~\ref{fig:RISPerf}. 

The results show that the proposed scheme of the AE-based RIS-aided communications outperforms the benchmark when the SNR exceeds a small threshold. Note that when the SNR is too small, then the channel for the RIS-aided communication is too weak to compensate the obstruction loss on the direct link. As the obstruction loss increases, the proposed scheme outperforms the benchmark scheme over a larger range of SNRs. For a given SER, Table~\ref{table:gain} shows the dB gain of the proposed scheme relative to the benchmark scheme with different obstruction losses ($l_o$ denotes the obstruction loss on the direct link.) As the target SER decreases or the obstruction loss on the direct link increases, the gain of the proposed scheme increases.

\begin{table}
	\caption{Performance gains of AE-based RIS-aided communications }
	\centering
	{\small
		\begin{tabular}{c|c|c|c}
			 & $l_o = 6$dB & $l_o = 7$dB & $l_o = 10$dB  \\ \hline \hline 
		    SER $= 10$e-$1$ & $2$dB Gain & $4$dB Gain & $10$dB Gain    \\ \hline 
		    SER $= 10$e-$2$ & $6$dB Gain & $8$dB Gain & $13$dB Gain    \\ \hline 
		    SER $= 10$e-$3$ & $8$dB Gain & $10$dB Gain & $16$dB Gain   \\ \hline 
		    SER $= 10$e-$4$ & $9$dB Gain & $11$dB Gain & $17$dB Gain    \\ \hline 
		    SER $= 10$e-$5$ & $10$dB Gain & $12$dB Gain & $18$dB Gain    \\ \hline 
		\end{tabular}
	}
	\label{table:gain}
\end{table}

Next, we evaluate the effect of the RIS beam selection on the end-to-end performance of SER at the receiver. We define the benchmark schemes, where the RIS selection is not for the best beam but for the top-$K$ beams, namely one of the best $K$ beam indices that lead to the smallest SERs is uniformly randomly selected. These types of benchmark schemes are typically used in practice to improve the accuracy of beam selection \cite{Nof}, when the sole objective is to select a good set of beams for the RIS.   

\begin{figure}
	\centerline{\includegraphics[width=1\linewidth]{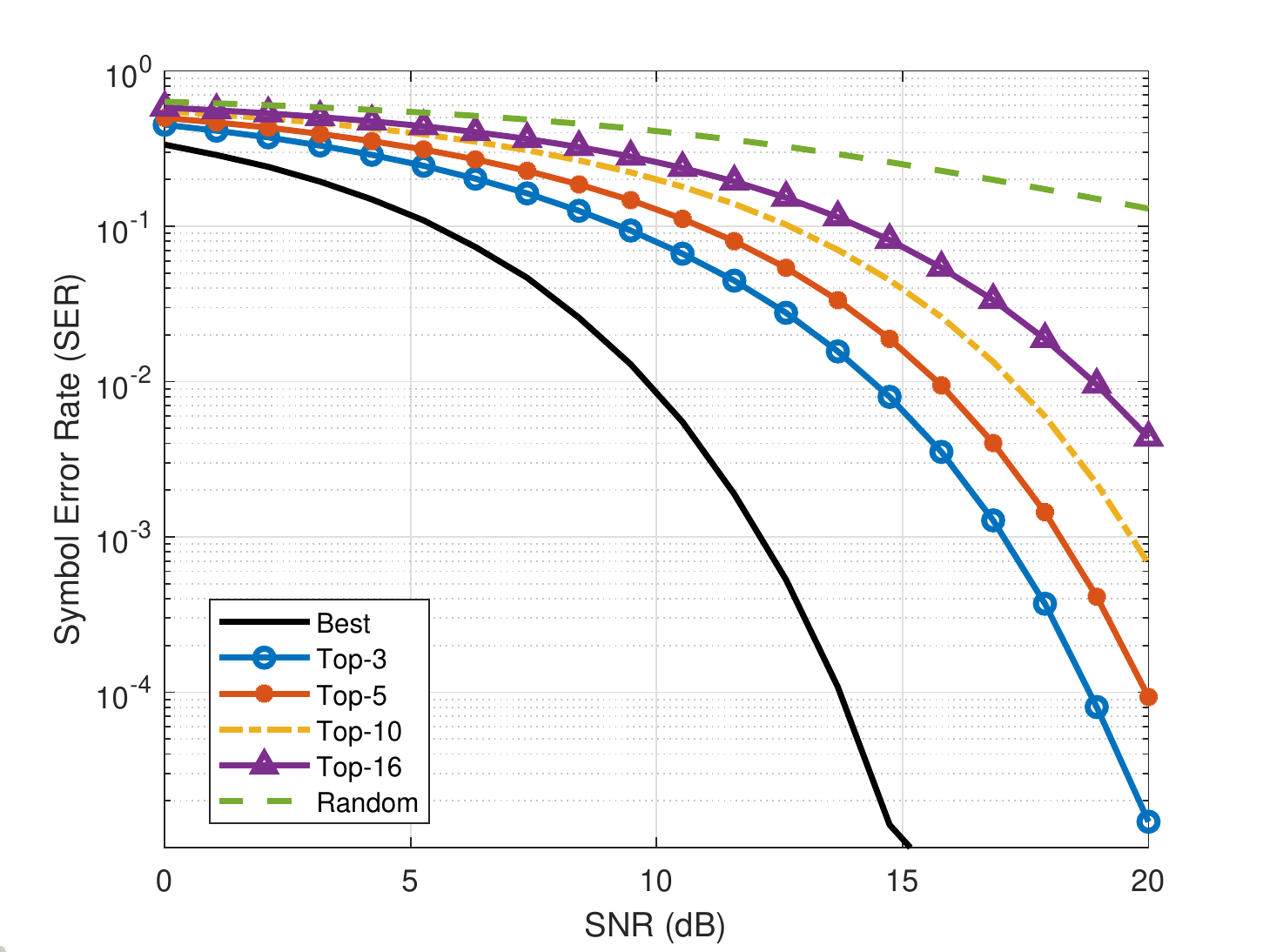}}
	\caption{SER vs. SNR performance when the best, random, top-3, top-5, top-10 and top-16 codebooks are selected.}
	\label{fig:TopRISPerf}
\end{figure}

Fig.~\ref{fig:TopRISPerf} evaluates the SER for benchmark schemes of top-$K$ beam selection, where $K = 1, 3, 5, 10, 16,$ and $32$. Note that $K=1$ is when the best beam is selected. The performance curve is the same as the RIS-Aided AE Link curve in Fig.~\ref{fig:RISPerf} when $K=1$. $K = 32$ is when any of the beams is randomly selected as there are total of $32$ beams considered. Note that the SER performance of the benchmark schemes drops as $K$ increases.

\section{Conclusion}\label{sec:Conclusion}
In this paper, we presented a design approach to optimize the transmitter-receiver pair along with the RIS beams by accounting for the channel data. Our approach uses an AE to model the transmitter and receiver operations via two DNNs. In addition, the selection of the RIS beam codebook is represented by another DNN. All these three DNNs are trained jointly with the objective of minimizing the SER at the receiver. In the presence of a physical obstruction between the transmitter and the receiver, we observe that this approach outperforms benchmarks including the use of direct communications without an RIS as well as the selection of the RIS codebook without the DNN. In particular, we show that it is possible to achieve significant reductions in the SER when the AE and RIS-aided communications are combined through joint training of the underlying DNNs. At the expense of increasing the complexity of the DNNs, it is expected that the performance gain will improve as more complex modulation types are used along with customized modulation and demodulation operations based on both channel and RIS beamforming capabilities. 



\end{document}